\documentclass[11pt]{article}
\usepackage[dvips]{graphicx}
\usepackage{enumitem}
\usepackage{amssymb}
\usepackage{color}
\usepackage{amsmath}
\usepackage{nicefrac}
\usepackage[left]{lineno}
\usepackage{hyperref}
\usepackage{adjustbox}
\usepackage{xspace}
\usepackage{multirow}
\usepackage{alphalph}
\usepackage[dvipsnames]{xcolor}

\definecolor{blu}{rgb}{0.,0.,1.}
\definecolor{red}{rgb}{1.,0.,0.}
\definecolor{burgundy}{rgb}{0.5, 0.0, 0.13}
\definecolor{crimsonred}{rgb}{0.6, 0.0, 0.0}
\definecolor{persianblue}{rgb}{0.11, 0.22, 0.73}
\definecolor{forestgreen}{rgb}{0.13,0.35,0.13}

\def\geant {\mbox{\textsc{Geant4}}\xspace}

\hypersetup{colorlinks, citecolor=crimsonred, linkcolor=persianblue, urlcolor=crimsonred}

\begin{document}
\centerline{\LARGE EUROPEAN ORGANIZATION FOR NUCLEAR RESEARCH}

%
\vspace{10mm} {\flushright{
CERN-EP-2022-243 \\
9 Nov 2022\\
\vspace{4mm}
Revised version:\\13 Jan 2023\\
}}
\vspace{-30mm}

%

%
\vspace{40mm}

\begin{center}
\boldmath
{\bf {\Large\boldmath{A search for the $K^+\to\mu^-\nu e^+e^+$ decay}}}
\unboldmath
\end{center}
\vspace{4mm}
\begin{center}
{\Large The NA62 Collaboration}\\
\end{center}

\begin{abstract}
A search for the $K^+\to\mu^-\nu e^+e^+$ decay, forbidden within the Standard Model by either lepton number or lepton flavour conservation depending on the flavour of the emitted neutrino, has been performed using the dataset collected by the NA62 experiment at CERN in 2016--2018. An upper limit of $8.1\times 10^{-11}$ is obtained for the decay branching fraction at 90\% CL, improving by a factor of 250 over the previous search.
\end{abstract}

\begin{center}
{\it Accepted for publication in Physics Letters B}
\end{center}

\newpage
\begin{center}
{\Large The NA62 Collaboration\renewcommand{\thefootnote}{\fnsymbol{footnote}}\footnotemark[1]\renewcommand{\thefootnote}{\arabic{footnote}}}\\
\end{center}
\begin{flushleft}
 E.~Cortina Gil\footnotemark[1],
 A.~Kleimenova\footnotemark[1]$^,$\renewcommand{\thefootnote}{\alphalph{\value{footnote}}}\footnotemark[1]\renewcommand{\thefootnote}{\arabic{footnote}},
E.~Minucci\footnotemark[1]$^,$\renewcommand{\thefootnote}{\alphalph{\value{footnote}}}\footnotemark[2]$^,$\footnotemark[3]\renewcommand{\thefootnote}{\arabic{footnote}},
 S.~Padolski\footnotemark[1]$^,$\renewcommand{\thefootnote}{\alphalph{\value{footnote}}}\footnotemark[4]\renewcommand{\thefootnote}{\arabic{footnote}},
 P.~Petrov\footnotemark[1],
 A.~Shaikhiev\footnotemark[1]$^,$\renewcommand{\thefootnote}{\alphalph{\value{footnote}}}\footnotemark[5]\renewcommand{\thefootnote}{\arabic{footnote}},
 R.~Volpe\footnotemark[1]$^,$\renewcommand{\thefootnote}{\alphalph{\value{footnote}}}\footnotemark[6]\renewcommand{\thefootnote}{\arabic{footnote}},  
 T.~Numao\footnotemark[2],
 Y.~Petrov\footnotemark[2],
 B.~Velghe\footnotemark[2],
 V.W.S.~Wong\footnotemark[2], 
 D.~Bryman\footnotemark[3]$^,$\renewcommand{\thefootnote}{\alphalph{\value{footnote}}}\footnotemark[7]\renewcommand{\thefootnote}{\arabic{footnote}},
 J.~Fu\footnotemark[3], 
 T.~Husek\footnotemark[4]$^,$\renewcommand{\thefootnote}{\alphalph{\value{footnote}}}\footnotemark[8]\renewcommand{\thefootnote}{\arabic{footnote}},
 J.~Jerhot\footnotemark[4]$^,$\renewcommand{\thefootnote}{\alphalph{\value{footnote}}}\footnotemark[9]\renewcommand{\thefootnote}{\arabic{footnote}},
 K.~Kampf\footnotemark[4],
 M.~Zamkovsky\footnotemark[4]$^,$\renewcommand{\thefootnote}{\alphalph{\value{footnote}}}\footnotemark[2]\renewcommand{\thefootnote}{\arabic{footnote}},
 R.~Aliberti\footnotemark[5]$^,$\renewcommand{\thefootnote}{\alphalph{\value{footnote}}}\footnotemark[10]\renewcommand{\thefootnote}{\arabic{footnote}},
 G.~Khoriauli\footnotemark[5]$^,$\renewcommand{\thefootnote}{\alphalph{\value{footnote}}}\footnotemark[11]\renewcommand{\thefootnote}{\arabic{footnote}},
 J.~Kunze\footnotemark[5],
 D.~Lomidze\footnotemark[5]$^,$\renewcommand{\thefootnote}{\alphalph{\value{footnote}}}\footnotemark[12]\renewcommand{\thefootnote}{\arabic{footnote}},
 L.~Peruzzo\footnotemark[5],
 M.~Vormstein\footnotemark[5],
 R.~Wanke\footnotemark[5],   
 P.~Dalpiaz\footnotemark[6],
 M.~Fiorini\footnotemark[6],
 I.~Neri\footnotemark[6],
 A.~Norton\footnotemark[6]$^,$\renewcommand{\thefootnote}{\alphalph{\value{footnote}}}\footnotemark[13]\renewcommand{\thefootnote}{\arabic{footnote}},
 F.~Petrucci\footnotemark[6],
 H.~Wahl\footnotemark[6]$^,$\renewcommand{\thefootnote}{\alphalph{\value{footnote}}}\footnotemark[14]\renewcommand{\thefootnote}{\arabic{footnote}},  
 A.~Cotta Ramusino\footnotemark[7],
 A.~Gianoli\footnotemark[7],  
 E.~Iacopini\footnotemark[8],
 G.~Latino\footnotemark[8],
 M.~Lenti\footnotemark[8],
 A.~Parenti\footnotemark[8], 
 A.~Bizzeti\footnotemark[9]$^,$\renewcommand{\thefootnote}{\alphalph{\value{footnote}}}\footnotemark[15]\renewcommand{\thefootnote}{\arabic{footnote}}, 
 F.~Bucci\footnotemark[9], 
 A.~Antonelli\footnotemark[10],
G.~Georgiev\footnotemark[10]$^,$\renewcommand{\thefootnote}{\alphalph{\value{footnote}}}\footnotemark[16]\renewcommand{\thefootnote}{\arabic{footnote}}, 
V.~Kozhuharov\footnotemark[10]$^,$\renewcommand{\thefootnote}{\alphalph{\value{footnote}}}\footnotemark[16]\renewcommand{\thefootnote}{\arabic{footnote}},
G.~Lanfranchi\footnotemark[10],
S.~Martellotti\footnotemark[10],
M.~Moulson\footnotemark[10],
T.~Spadaro\footnotemark[10],
G.~Tinti\footnotemark[10], 
 F.~Ambrosino\footnotemark[11],
 T.~Capussela\footnotemark[11],
 M.~Corvino\footnotemark[11]$^,$\renewcommand{\thefootnote}{\alphalph{\value{footnote}}}\footnotemark[2]\renewcommand{\thefootnote}{\arabic{footnote}}, 
 D.~Di Filippo\footnotemark[11],
 R.~Fiorenza\footnotemark[11]$^,$\renewcommand{\thefootnote}{\alphalph{\value{footnote}}}\footnotemark[17]\renewcommand{\thefootnote}{\arabic{footnote}}, 
 P.~Massarotti\footnotemark[11],
 M.~Mirra\footnotemark[11],
 M.~Napolitano\footnotemark[11],
 G.~Saracino\footnotemark[11],  
 G.~Anzivino\footnotemark[12],
 F.~Brizioli\footnotemark[12]$^,$\renewcommand{\thefootnote}{\alphalph{\value{footnote}}}\footnotemark[2]\renewcommand{\thefootnote}{\arabic{footnote}}, 
 E.~Imbergamo\footnotemark[12],
 R.~Lollini\footnotemark[12],
 R.~Piandani\footnotemark[12]$^,$\renewcommand{\thefootnote}{\alphalph{\value{footnote}}}\footnotemark[18]\renewcommand{\thefootnote}{\arabic{footnote}}, 
 C.~Santoni\footnotemark[12],  
 M.~Barbanera\footnotemark[13],
 P.~Cenci\footnotemark[13],
 B.~Checcucci\footnotemark[13],
 P.~Lubrano\footnotemark[13],
 M.~Lupi\footnotemark[13]$^,$\renewcommand{\thefootnote}{\alphalph{\value{footnote}}}\footnotemark[19]\renewcommand{\thefootnote}{\arabic{footnote}}, 
 M.~Pepe\footnotemark[13],
 M.~Piccini\footnotemark[13],  
{
 F.~Costantini\footnotemark[14],
L.~Di Lella\footnotemark[14]$^,$\renewcommand{\thefootnote}{\alphalph{\value{footnote}}}\footnotemark[14]\renewcommand{\thefootnote}{\arabic{footnote}}, 
 N.~Doble\footnotemark[14]$^,$\renewcommand{\thefootnote}{\alphalph{\value{footnote}}}\footnotemark[14]\renewcommand{\thefootnote}{\arabic{footnote}}, 
 M.~Giorgi\footnotemark[14],
 S.~Giudici\footnotemark[14],
 G.~Lamanna\footnotemark[14],
 E.~Lari\footnotemark[14],
 E.~Pedreschi\footnotemark[14],
 M.~Sozzi\footnotemark[14],  
 C.~Cerri\footnotemark[15],
 R.~Fantechi\footnotemark[15],
 L.~Pontisso\footnotemark[15]$^,$\renewcommand{\thefootnote}{\alphalph{\value{footnote}}}\footnotemark[20]\renewcommand{\thefootnote}{\arabic{footnote}}, 
 F.~Spinella\footnotemark[15],  
 I.~Mannelli\footnotemark[16],   
 G.~D'Agostini\footnotemark[17], 
 M.~Raggi\footnotemark[17],  
 A.~Biagioni\footnotemark[18], 
 P.~Cretaro\footnotemark[18], 
 O.~Frezza\footnotemark[18], 
 E.~Leonardi\footnotemark[18], 
 A.~Lonardo\footnotemark[18], 
 M.~Turisini\footnotemark[18], 
 P.~Valente\footnotemark[18], 
 P.~Vicini\footnotemark[18],  
 R.~Ammendola\footnotemark[19], 
 V.~Bonaiuto\footnotemark[19]$^,$\renewcommand{\thefootnote}{\alphalph{\value{footnote}}}\footnotemark[21]\renewcommand{\thefootnote}{\arabic{footnote}}, 
 A.~Fucci\footnotemark[19], 
 A.~Salamon\footnotemark[19], 
 F.~Sargeni\footnotemark[19]$^,$\renewcommand{\thefootnote}{\alphalph{\value{footnote}}}\footnotemark[22]\renewcommand{\thefootnote}{\arabic{footnote}},   
 R.~Arcidiacono\footnotemark[20]$^,$\renewcommand{\thefootnote}{\alphalph{\value{footnote}}}\footnotemark[23]\renewcommand{\thefootnote}{\arabic{footnote}}, 
 B.~Bloch-Devaux\footnotemark[20],
 M.~Boretto\footnotemark[20]$^,$\renewcommand{\thefootnote}{\alphalph{\value{footnote}}}\footnotemark[2]\renewcommand{\thefootnote}{\arabic{footnote}}, 
 E.~Menichetti\footnotemark[20],
 E.~Migliore\footnotemark[20],
 D.~Soldi\footnotemark[20],   
 C.~Biino\footnotemark[21],
 A.~Filippi\footnotemark[21],
 F.~Marchetto\footnotemark[21],  
 J.~Engelfried\footnotemark[22],
 N.~Estrada-Tristan\footnotemark[22]$^,$\renewcommand{\thefootnote}{\alphalph{\value{footnote}}}\footnotemark[24]\renewcommand{\thefootnote}{\arabic{footnote}},  
 A.M.~Bragadireanu\footnotemark[23],
 S.A.~Ghinescu\footnotemark[23],
 O.E.~Hutanu\footnotemark[23], 
 A.~Baeva\footnotemark[24],
 D.~Baigarashev\footnotemark[24]$^,$\renewcommand{\thefootnote}{\alphalph{\value{footnote}}}\footnotemark[25]\renewcommand{\thefootnote}{\arabic{footnote}}, 
 D.~Emelyanov\footnotemark[24],
 T.~Enik\footnotemark[24],
 V.~Falaleev\footnotemark[24]$^,$\renewcommand{\thefootnote}{\alphalph{\value{footnote}}}\footnotemark[26]\renewcommand{\thefootnote}{\arabic{footnote}}, 
 V.~Kekelidze\footnotemark[24],
 A.~Korotkova\footnotemark[24],
 L.~Litov\footnotemark[24]$^,$\renewcommand{\thefootnote}{\alphalph{\value{footnote}}}\footnotemark[16]\renewcommand{\thefootnote}{\arabic{footnote}}, 
 D.~Madigozhin\footnotemark[24],
 M.~Misheva\footnotemark[24]$^,$\renewcommand{\thefootnote}{\alphalph{\value{footnote}}}\footnotemark[27]\renewcommand{\thefootnote}{\arabic{footnote}}, 
 N.~Molokanova\footnotemark[24],
 S.~Movchan\footnotemark[24],
 I.~Polenkevich\footnotemark[24],
 Yu.~Potrebenikov\footnotemark[24],
 S.~Shkarovskiy\footnotemark[24],
 A.~Zinchenko\footnotemark[24]$^,$\renewcommand{\thefootnote}{\fnsymbol{footnote}}\footnotemark[2]\renewcommand{\thefootnote}{\arabic{footnote}},  
 S.~Fedotov\footnotemark[25],
 E.~Gushchin\footnotemark[25],
 A.~Khotyantsev\footnotemark[25],
 Y.~Kudenko\footnotemark[25]$^,$\renewcommand{\thefootnote}{\alphalph{\value{footnote}}}\footnotemark[28]\renewcommand{\thefootnote}{\arabic{footnote}},  
 V.~Kurochka\footnotemark[25],
 M.~Medvedeva\footnotemark[25],
 A.~Mefodev \footnotemark[25], 
 S.~Kholodenko\footnotemark[26],
 V.~Kurshetsov\footnotemark[26],
 V.~Obraztsov\footnotemark[26],
 A.~Ostankov\footnotemark[26]$^,$\renewcommand{\thefootnote}{\fnsymbol{footnote}}\footnotemark[2]\renewcommand{\thefootnote}{\arabic{footnote}},
 V.~Semenov\footnotemark[26]$^,$\renewcommand{\thefootnote}{\fnsymbol{footnote}}\footnotemark[2]\renewcommand{\thefootnote}{\arabic{footnote}},
 V.~Sugonyaev\footnotemark[26],
 O.~Yushchenko\footnotemark[26],  
 L.~Bician\footnotemark[27]$^,$\renewcommand{\thefootnote}{\alphalph{\value{footnote}}}\footnotemark[29]\renewcommand{\thefootnote}{\arabic{footnote}},  
 T.~Blazek\footnotemark[27],
 V.~Cerny\footnotemark[27],
 Z.~Kucerova\footnotemark[27],  
 J.~Bernhard\footnotemark[28],
 A.~Ceccucci\footnotemark[28],
 H.~Danielsson\footnotemark[28],
 N.~De Simone\footnotemark[28]$^,$\renewcommand{\thefootnote}{\alphalph{\value{footnote}}}\footnotemark[30]\renewcommand{\thefootnote}{\arabic{footnote}},
 F.~Duval\footnotemark[28],
 B.~D\"obrich\footnotemark[28],
 L.~Federici\footnotemark[28],
 E.~Gamberini\footnotemark[28],
 L.~Gatignon\footnotemark[28]$^,$\renewcommand{\thefootnote}{\alphalph{\value{footnote}}}\footnotemark[31]\renewcommand{\thefootnote}{\arabic{footnote}},
 R.~Guida\footnotemark[28],
 F.~Hahn\footnotemark[28]$^,$\renewcommand{\thefootnote}{\fnsymbol{footnote}}\footnotemark[2]\renewcommand{\thefootnote}{\arabic{footnote}},
 E.B.~Holzer\footnotemark[28],
 B.~Jenninger\footnotemark[28],
 M.~Koval\footnotemark[28]$^,$\renewcommand{\thefootnote}{\alphalph{\value{footnote}}}\footnotemark[29]\renewcommand{\thefootnote}{\arabic{footnote}},
 P.~Laycock\footnotemark[28]$^,$\renewcommand{\thefootnote}{\alphalph{\value{footnote}}}\footnotemark[4]\renewcommand{\thefootnote}{\arabic{footnote}},
 G.~Lehmann Miotto\footnotemark[28],
 P.~Lichard\footnotemark[28],
 A.~Mapelli\footnotemark[28],
 R.~Marchevski\footnotemark[28]$^,$\renewcommand{\thefootnote}{\alphalph{\value{footnote}}}\footnotemark[32]\renewcommand{\thefootnote}{\arabic{footnote}},
 K.~Massri\footnotemark[28],
 M.~Noy\footnotemark[28],
 V.~Palladino\footnotemark[28],
 M.~Perrin-Terrin\footnotemark[28]$^,$\renewcommand{\thefootnote}{\alphalph{\value{footnote}}}\footnotemark[33]$^,$\footnotemark[34]\renewcommand{\thefootnote}{\arabic{footnote}},
 J.~Pinzino\footnotemark[28]$^,$\renewcommand{\thefootnote}{\alphalph{\value{footnote}}}\footnotemark[35]\renewcommand{\thefootnote}{\arabic{footnote}}\renewcommand{\thefootnote}{\arabic{footnote}},
 V.~Ryjov\footnotemark[28],
 S.~Schuchmann\footnotemark[28],
 S.~Venditti\footnotemark[28],  
 T.~Bache\footnotemark[29],
 M.B.~Brunetti\footnotemark[29]$^,$\renewcommand{\thefootnote}{\alphalph{\value{footnote}}}\footnotemark[36]\renewcommand{\thefootnote}{\arabic{footnote}},
 V.~Duk\footnotemark[29]$^,$\renewcommand{\thefootnote}{\alphalph{\value{footnote}}}\footnotemark[6]\renewcommand{\thefootnote}{\arabic{footnote}},
 V.~Fascianelli\footnotemark[29]$^,$\renewcommand{\thefootnote}{\alphalph{\value{footnote}}}\footnotemark[37]\renewcommand{\thefootnote}{\arabic{footnote}},
 J.R.~Fry\footnotemark[29],
 F.~Gonnella\footnotemark[29],
 E.~Goudzovski\footnotemark[29]$^,$\renewcommand{\thefootnote}{\fnsymbol{footnote}}\footnotemark[1]\renewcommand{\thefootnote}{\arabic{footnote}},
 J.~Henshaw\footnotemark[29],
 L.~Iacobuzio\footnotemark[29],
 C.~Lazzeroni\footnotemark[29],
 N.~Lurkin\footnotemark[29]$^,$\renewcommand{\thefootnote}{\alphalph{\value{footnote}}}\footnotemark[9]\renewcommand{\thefootnote}{\arabic{footnote}},
 F.~Newson\footnotemark[29],
 C.~Parkinson\footnotemark[29], 
 A.~Romano\footnotemark[29], 
 A.~Sergi\footnotemark[29]$^,$\renewcommand{\thefootnote}{\alphalph{\value{footnote}}}\footnotemark[38]\renewcommand{\thefootnote}{\arabic{footnote}},
 A.~Sturgess\footnotemark[29],
 J.~Swallow\footnotemark[29]$^,$\renewcommand{\thefootnote}{\alphalph{\value{footnote}}}\footnotemark[2]\renewcommand{\thefootnote}{\arabic{footnote}},
 A.~Tomczak\footnotemark[29],  
 H.~Heath\footnotemark[30],
 R.~Page\footnotemark[30],
 S.~Trilov\footnotemark[30], 
 B.~Angelucci\footnotemark[31],
 D.~Britton\footnotemark[31],
 C.~Graham\footnotemark[31],
 D.~Protopopescu\footnotemark[31],  
 J.~Carmignani\footnotemark[32]$^,$\renewcommand{\thefootnote}{\alphalph{\value{footnote}}}\footnotemark[39]\renewcommand{\thefootnote}{\arabic{footnote}}, 
 J.B.~Dainton\footnotemark[32],
 R.W.L.~Jones\footnotemark[32],
 G.~Ruggiero\footnotemark[32]$^,$\renewcommand{\thefootnote}{\alphalph{\value{footnote}}}\footnotemark[40]\renewcommand{\thefootnote}{\arabic{footnote}},   
 L.~Fulton\footnotemark[33],
 D.~Hutchcroft\footnotemark[33],
 E.~Maurice\footnotemark[33]$^,$\renewcommand{\thefootnote}{\alphalph{\value{footnote}}}\footnotemark[41]\renewcommand{\thefootnote}{\arabic{footnote}},  
 B.~Wrona\footnotemark[33],  
 A.~Conovaloff\footnotemark[34],
 P.~Cooper\footnotemark[34],
 D.~Coward\footnotemark[34]$^,$\renewcommand{\thefootnote}{\alphalph{\value{footnote}}}\footnotemark[42]\renewcommand{\thefootnote}{\arabic{footnote}},  
 P.~Rubin\footnotemark[34]  
} \end{flushleft}
\newlength{\basefootnotesep}
\setlength{\basefootnotesep}{\footnotesep}
\begin{flushleft}
\setcounter{footnote}{0}
\renewcommand{\thefootnote}{\fnsymbol{footnote}}
\footnotetext[1]{Corresponding author:  E.~Goudzovski, email: evgueni.goudzovski@cern.ch}
\footnotetext[2]{Deceased}
\renewcommand{\thefootnote}{\arabic{footnote}}
$^{1}$ 
Universit\'e Catholique de Louvain, B-1348 Louvain-La-Neuve, Belgium \\
%
$^{2}$ 
TRIUMF, Vancouver, British Columbia, V6T 2A3, Canada \\
%
$^{3}$
University of British Columbia, Vancouver, British Columbia, V6T 1Z4, Canada \\
$^{4}$
Charles University, 116 36 Prague 1, Czech Republic \\
%
$^{5}$
Institut f\"ur Physik and PRISMA Cluster of Excellence, Universit\"at Mainz, D-55099 Mainz, Germany \\
%
$^{6}$
Dipartimento di Fisica e Scienze della Terra dell'Universit\`a e INFN, Sezione di Ferrara, I-44122 Ferrara, Italy \\
%
$^{7}$
INFN, Sezione di Ferrara, I-44122 Ferrara, Italy \\
%
$^{8}$
Dipartimento di Fisica e Astronomia dell'Universit\`a e INFN, Sezione di Firenze, I-50019 Sesto Fiorentino, Italy \\
%
$^{9}$
INFN, Sezione di Firenze, I-50019 Sesto Fiorentino, Italy \\
%
$^{10}$
Laboratori Nazionali di Frascati, I-00044 Frascati, Italy \\
%
$^{11}$
Dipartimento di Fisica ``Ettore Pancini'' e INFN, Sezione di Napoli, I-80126 Napoli, Italy \\
%
$^{12}$
Dipartimento di Fisica e Geologia dell'Universit\`a e INFN, Sezione di Perugia, I-06100 Perugia, Italy \\
%
$^{13}$
INFN, Sezione di Perugia, I-06100 Perugia, Italy \\
%
$^{14}$
Dipartimento di Fisica dell'Universit\`a e INFN, Sezione di Pisa, I-56100 Pisa, Italy \\
%
$^{15}$
INFN, Sezione di Pisa, I-56100 Pisa, Italy \\
%
$^{16}$
Scuola Normale Superiore e INFN, Sezione di Pisa, I-56100 Pisa, Italy \\
%
$^{17}$
Dipartimento di Fisica, Sapienza Universit\`a di Roma e INFN, Sezione di Roma I, I-00185 Roma, Italy \\
$^{18}$
INFN, Sezione di Roma I, I-00185 Roma, Italy \\
%
$^{19}$
INFN, Sezione di Roma Tor Vergata, I-00133 Roma, Italy \\
%
$^{20}$
Dipartimento di Fisica dell'Universit\`a e INFN, Sezione di Torino, I-10125 Torino, Italy \\
%
$^{21}$
INFN, Sezione di Torino, I-10125 Torino, Italy \\
%
$^{22}$
Instituto de F\'isica, Universidad Aut\'onoma de San Luis Potos\'i, 78240 San Luis Potos\'i, Mexico \\
%
$^{23}$
Horia Hulubei National Institute for R\&D in Physics and Nuclear Engineering, 077125 Bucharest-Magurele, Romania \\
%
$^{24}$
Joint Institute for Nuclear Research, 141980 Dubna (MO), Russia \\
%
$^{25}$
Institute for Nuclear Research of the Russian Academy of Sciences, 117312 Moscow, Russia \\
%
$^{26}$
Institute for High Energy Physics of the Russian Federation, State Research Center ``Kurchatov Institute", 142281 Protvino (MO), Russia \\
%
$^{27}$
Faculty of Mathematics, Physics and Informatics, Comenius University, 842 48, Bratislava, Slovakia \\
%
$^{28}$
CERN,  European Organization for Nuclear Research, CH-1211 Geneva 23, Switzerland \\
%
$^{29}$
School of Physics and Astronomy, University of Birmingham, Edgbaston, Birmingham, B15 2TT, UK \\
%
$^{30}$
School of Physics, University of Bristol, Bristol, BS8 1TH, UK \\
%
$^{31}$
School of Physics and Astronomy, University of Glasgow, Glasgow, G12 8QQ, UK \\
%
$^{32}$
Faculty of Science and Technology, University of Lancaster, Lancaster, LA1 4YW, UK \\
%
$^{33}$
School of Physical Sciences, University of Liverpool, Liverpool, L69 7ZE, UK \\
%
$^{34}$
Physics and Astronomy Department, George Mason University, Fairfax, VA 22030, USA
%
\end{flushleft}
\renewcommand{\thefootnote}{\alphalph{\value{footnote}}}
\footnotesize
{$^{\alphalph{1}}$Present address: Faculty of Mathematics, Physics and Informatics, Comenius University, 842 48, Bratislava, Slovakia \\
%
$^{\alphalph{2}}$Present address: CERN,  European Organization for Nuclear Research, CH-1211 Geneva 23, Switzerland \\
%
$^{\alphalph{3}}$Also at Laboratori Nazionali di Frascati, I-00044 Frascati, Italy \\
%
$^{\alphalph{4}}$Present address: Brookhaven National Laboratory, Upton, NY 11973, USA \\
%
$^{\alphalph{5}}$Present address: School of Physics and Astronomy, University of Birmingham, Birmingham, B15 2TT, UK \\
%
$^{\alphalph{6}}$Present address: INFN, Sezione di Perugia, I-06100 Perugia, Italy \\
%
$^{\alphalph{7}}$Also at TRIUMF, Vancouver, British Columbia, V6T 2A3, Canada \\
%
$^{\alphalph{8}}$Present address: Department of Astronomy and Theoretical Physics, Lund University, Lund, SE 223-62, Sweden \\
%
$^{\alphalph{9}}$Present address: Universit\'e Catholique de Louvain, B-1348 Louvain-La-Neuve, Belgium \\
%
$^{\alphalph{10}}$Present address: Institut f\"ur Kernphysik and Helmholtz Institute Mainz, Universit\"at Mainz, Mainz, D-55099, Germany \\
%
$^{\alphalph{11}}$Present address: Universit\"at W\"urzburg, D-97070 W\"urzburg, Germany \\
%
$^{\alphalph{12}}$Present address: European XFEL GmbH, D-22761 Hamburg, Germany \\
%
$^{\alphalph{13}}$Present address: University of Glasgow, Glasgow, G12 8QQ, UK \\
%
$^{\alphalph{14}}$Present address: Institut f\"ur Physik and PRISMA Cluster of Excellence, Universit\"at Mainz, D-55099 Mainz, Germany \\
%
$^{\alphalph{15}}$Also at Dipartimento di Scienze Fisiche, Informatiche e Matematiche, Universit\`a di Modena e Reggio Emilia, I-41125 Modena, Italy \\
%
$^{\alphalph{16}}$Also at Faculty of Physics, University of Sofia, BG-1164 Sofia, Bulgaria \\
%
$^{\alphalph{17}}$Present address: Scuola Superiore Meridionale e INFN, Sezione di Napoli, I-80138 Napoli, Italy \\
%
$^{\alphalph{18}}$Present address: Instituto de F\'isica, Universidad Aut\'onoma de San Luis Potos\'i, 78240 San Luis Potos\'i, Mexico \\
%
$^{\alphalph{19}}$Present address: Institut am Fachbereich Informatik und Mathematik, Goethe Universit\"at, D-60323 Frankfurt am Main, Germany \\
%
$^{\alphalph{20}}$Present address: INFN, Sezione di Roma I, I-00185 Roma, Italy \\
%
$^{\alphalph{21}}$Also at Department of Industrial Engineering, University of Roma Tor Vergata, I-00173 Roma, Italy \\
%
$^{\alphalph{22}}$Also at Department of Electronic Engineering, University of Roma Tor Vergata, I-00173 Roma, Italy \\
%
$^{\alphalph{23}}$Also at Universit\`a degli Studi del Piemonte Orientale, I-13100 Vercelli, Italy \\
%
$^{\alphalph{24}}$Also at Universidad de Guanajuato, 36000 Guanajuato, Mexico \\
%
$^{\alphalph{25}}$Also at L.N. Gumilyov Eurasian National University, 010000 Nur-Sultan, Kazakhstan \\
%
$^{\alphalph{26}}$Also at Institute for Nuclear Research of the Russian Academy of Sciences, 117312 Moscow, Russia \\
%
$^{\alphalph{27}}$Present address: Institute of Nuclear Research and Nuclear Energy of Bulgarian Academy of Science (INRNE-BAS), BG-1784 Sofia, Bulgaria \\
%
$^{\alphalph{28}}$Also at National Research Nuclear University (MEPhI), 115409 Moscow and Moscow Institute of Physics and Technology, 141701 Moscow region, Moscow, Russia \\
%
$^{\alphalph{29}}$Present address: Charles University, 116 36 Prague 1, Czech Republic \\
%
$^{\alphalph{30}}$Present address: DESY, D-15738 Zeuthen, Germany \\
%
$^{\alphalph{31}}$Present address: University of Lancaster, Lancaster, LA1 4YW, UK \\
%
$^{\alphalph{32}}$Present address: Weizmann Institute, Rehovot, 76100, Israel \\
%
$^{\alphalph{33}}$Present address: Aix Marseille University, CNRS/IN2P3, CPPM, F-13288, Marseille, France \\
%
$^{\alphalph{34}}$Also at Universit\'e Catholique de Louvain, B-1348 Louvain-La-Neuve, Belgium \\
%
$^{\alphalph{35}}$Present address: INFN, Sezione di Pisa, I-56100 Pisa, Italy \\
%
$^{\alphalph{36}}$Present address: Department of Physics, University of Warwick, Coventry, CV4 7AL, UK \\
%
$^{\alphalph{37}}$Present address: Center for theoretical neuroscience, Columbia University, New York, NY 10027, USA \\
%
$^{\alphalph{38}}$Present address: Dipartimento di Fisica dell'Universit\`a e INFN, Sezione di Genova, I-16146 Genova, Italy \\
%
$^{\alphalph{39}}$Present address: University of Liverpool, Liverpool, L69 7ZE, UK \\
%
$^{\alphalph{40}}$Present address: Dipartimento di Fisica e Astronomia dell'Universit\`a e INFN, Sezione di Firenze, I-50019 Sesto Fiorentino, Italy \\
%
$^{\alphalph{41}}$Present address: Laboratoire Leprince Ringuet, F-91120 Palaiseau, France \\
%
$^{\alphalph{42}}$Also at SLAC National Accelerator Laboratory, Stanford University, Menlo Park, CA 94025, USA
}
\normalsize
\clearpage
\clearpage



\newpage

\section*{Introduction}
\vspace{-1mm}

In the Standard Model (SM), neutrinos are strictly massless due to the absence of right-handed chiral states. The discovery of neutrino oscillations has demonstrated non-zero neutrino mass, which makes the experimental discrimination between the Dirac and Majorana neutrino possible in principle. Strong evidence for the Majorana nature of the neutrino would be provided by the observation of lepton number violating (LNV) processes, including kaon decays~\cite{li00,atre05,atre09,ab18}.
Furthermore, lepton flavour violating (LFV) kaon decays are expected in new physics models involving flavour violating ALPs and $Z'$ particles~\cite{alp2020,review}.

The NA62 experiment at CERN collected a large dataset of $K^+$ decays to lepton pairs in 2016--2018, using dedicated trigger lines. This dataset has been analysed to establish stringent upper limits of the branching ratios of the LNV decays $K^+\to\pi^-(\pi^0)e^+e^+$~\cite{co22}, $K^+\to\pi^-\mu^+\mu^+$~\cite{co19} and $K^+\to\pi^-\mu^+e^+$~\cite{co21}, as well as LFV decays $K^+\to\pi^+\mu^-e^+$ and $\pi^0\to\mu^-e^+$~\cite{co21}.

The $K^+\to\mu^-\nu e^+e^+$ decay is forbidden in the SM by either LN or LF conservation, depending on the flavour of the emitted neutrino. Experimentally, the current upper limit of the decay branching fraction is $2.1\times 10^{-8}$ at 90\% CL~\cite{di76,pdg}. In the context of the above programme, a new search for the $K^+\to\mu^-\nu e^+e^+$ decay with the NA62 2016--2018 dataset is reported here.


\vspace{-1.2mm}
\section{Beam, detector and data sample}
\label{sec:detector}
\vspace{-1mm}

\begin{figure}[t]
\begin{center}
\resizebox{\textwidth}{!}{\includegraphics{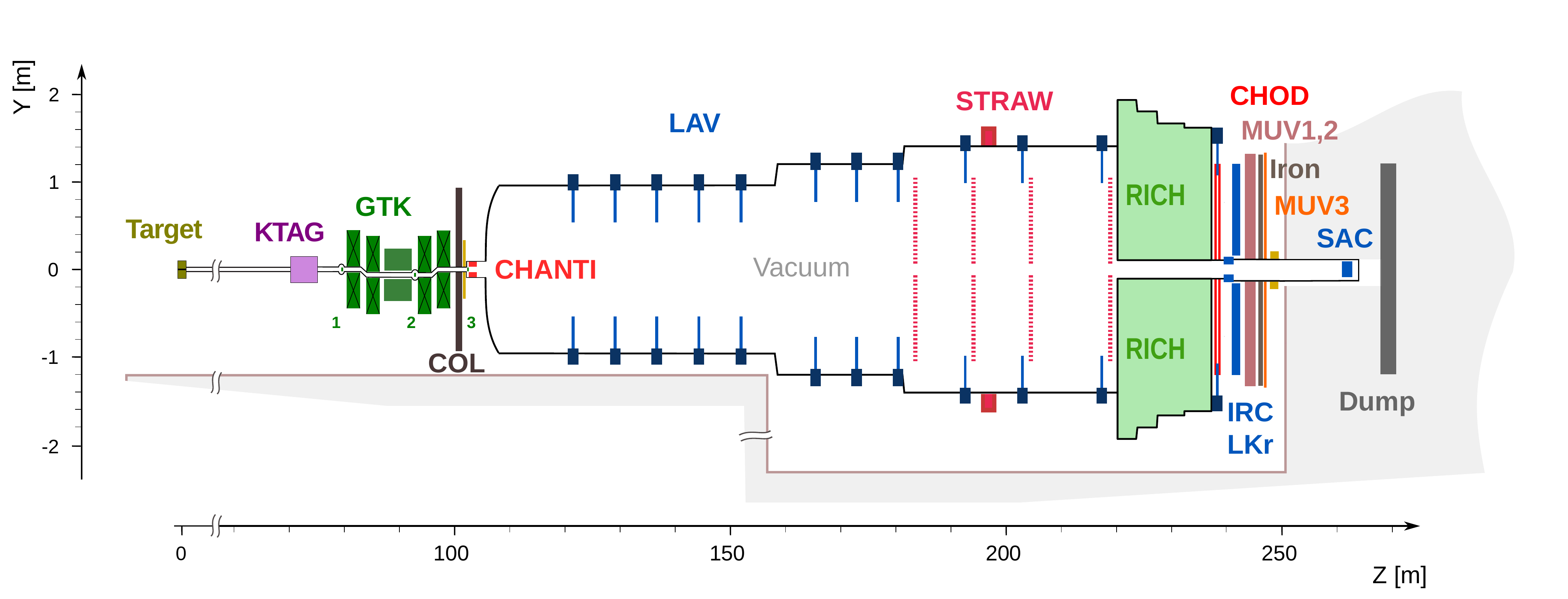}}
\put(-343,68){\scriptsize\color{OliveGreen}\rotatebox{90}{\textbf{\textsf{SCR}}}}
\put(-171.5,46){\scriptsize\color{red}{\textbf{\textsf{M}}}}
\put(-319.5,103){\tiny\color{YellowOrange}\rotatebox{90}{\textbf{\textsf{TRIM5}}}}
\end{center}
\vspace{-14mm}
\caption{Schematic side view of the NA62 beamline and detector.}
\label{fig:detector}
\end{figure}

The layout of the NA62 beamline and detector~\cite{na62-detector} is shown schematically in Fig.~\ref{fig:detector}. An unseparated secondary beam of $\pi^+$ (70\%), protons (23\%) and $K^+$ (6\%) is created by directing 400~GeV/$c$ protons extracted from the CERN SPS onto a beryllium target in spills of 3~s effective duration.
The beam central momentum is 75~GeV/$c$, with a momentum spread of 1\% (rms).

Beam kaons are tagged with a time resolution of 70~ps by a differential Cherenkov counter (KTAG), which uses nitrogen gas at 1.75~bar pressure contained in a 5~m long vessel as radiator. Beam particle positions, momenta and times (to better than 100~ps resolution) are measured by a silicon pixel spectrometer consisting of three stations (GTK1,2,3) and four dipole magnets forming an achromat. A toroidal muon sweeper (scraper, SCR) is installed between GTK1 and GTK2. A 1.2~m thick steel collimator (COL) with a $76\times40$~mm$^2$ central aperture and $1.7\times1.8$~m$^2$ outer dimensions is placed upstream of GTK3 to absorb hadrons from upstream $K^+$ decays; a variable-aperture collimator of \mbox{$0.15\times0.15$~m$^2$} outer dimensions was used up to early 2018. Inelastic interactions of beam particles in GTK3 are detected by an array of scintillator hodoscopes (CHANTI). A dipole magnet (TRIM5) providing a 90 MeV/c horizontal momentum kick is located in front of GTK3. The beam is delivered into a vacuum tank evacuated to a pressure of $10^{-6}$~mbar, which contains a 75~m long fiducial volume (FV) starting 2.6~m downstream of GTK3. The beam angular spread at the FV entrance is 0.11~mrad (rms) in both horizontal and vertical planes. Downstream of the FV, undecayed beam particles continue their path in vacuum.

Momenta of charged particles produced in $K^+$ decays in the FV are measured by a magnetic spectrometer (STRAW) located in the vacuum tank downstream of the FV. The spectrometer consists of four tracking chambers made of straw tubes, and a dipole magnet (M) located between the second and third chambers that provides a horizontal momentum kick of 270~MeV/$c$ in a direction opposite to that produced by TRIM5. The momentum resolution is $\sigma_p/p = (0.30\oplus 0.005\cdot p)\%$, with the momentum $p$ expressed in GeV/$c$.

%
A ring-imaging Cherenkov detector (RICH) consisting of a 17.5~m long vessel filled with neon at atmospheric pressure (with a Cherenkov threshold of 12.5~GeV/$c$ for pions) provides particle identification, charged particle time measurements with a typical resolution of 70~ps, and the trigger time.
Two scintillator hodoscopes (CHOD), which include a matrix of tiles and two planes of slabs arranged in four quadrants located downstream of the RICH, provide trigger signals and time measurements with 200~ps precision.

\newpage

A $27X_0$ thick quasi-homogeneous liquid-krypton (LKr) electromagnetic calorimeter is used for particle identification and photon detection. The calorimeter has an active volume of 7~m$^3$, segmented in the transverse direction into 13248 projective cells of
$2\times 2$~cm$^2$ size, and provides an energy resolution $\sigma_E/E=(4.8/\sqrt{E}\oplus11/E\oplus0.9)\%$, with $E$ expressed in GeV. To achieve hermetic acceptance for photons emitted in $K^+$ decays in the FV at angles up to 50~mrad from the beam axis, the LKr calorimeter is complemented by annular lead glass detectors (LAV) installed in 12~positions inside and downstream of the vacuum tank, and two lead/scintillator sampling calorimeters (IRC, SAC) located close to the beam axis. An iron/scintillator sampling hadronic calorimeter formed of two modules (MUV1,2) and a muon detector consisting of 148~scintillator tiles located behind an 80~cm thick iron wall (MUV3) are used for particle identification.

%
%

The data sample analysed is obtained from $0.89\times 10^6$ SPS spills recorded in 2016--2018, with the typical beam intensity increasing over time from \mbox{$1.3\times 10^{12}$} to \mbox{$2.2\times 10^{12}$} protons per spill. The latter value corresponds to a 500~MHz mean instantaneous beam particle rate at the FV entrance, and a 3.7~MHz mean $K^+$ decay rate in the FV. The main NA62 trigger is dedicated to the collection of very rare $K^+\to\pi^+\nu\bar\nu$ decays~\cite{pinn}. The present analysis is based on the dedicated multi-track (MT), electron multi-track ($e$MT) and muon multi-track ($\mu$MT) trigger lines operating concurrently with the main trigger line~\cite{na62-trigger,am19}, downscaled typically by factors $D_{\rm MT}=100$, $D_{e\rm MT}=8$ and $D_{\mu\rm MT}=8$. The downscaling factors were varied during data-taking to accommodate the increasing beam intensity. The low-level (L0) hardware trigger is based on RICH signal multiplicity and coincidence of signals in two opposite CHOD quadrants. The $\mu$MT ($e$MT) line requires a 10~(20)~GeV energy deposit in the LKr calorimeter.
The $\mu$MT line requires a signal in an outer tile of the MUV3 detector (i.e. one of the 140 tiles not adjacent to the beam pipe). The high-level (L1) software trigger involves beam $K^+$ identification by the KTAG, reconstruction of a negatively-charged STRAW track, and fewer than three in-time signals in LAV stations 2--11 (in the $\mu$MT trigger line only). For signal-like samples characterised by an LKr energy deposit well above 20~GeV, the measured inefficiencies of the CHOD (STRAW) trigger conditions are typically at the 1\%~(5\%) level, while those of the RICH, MUV3, KTAG and LKr conditions are of ${\cal O}(10^{-3})$.

Monte Carlo (MC) simulations of particle interactions with the detector and its response are performed using a software package based on the~\geant toolkit~\cite{geant4}. In addition, accidental activity is simulated and the response of the trigger lines is emulated.


\section{Event selection}
\label{sec:selection}
\vspace{-1mm}

The rate of the possible signal decay $K^+\to\mu^-\nu e^+e^+$ (denoted $K_{\mu\nu ee}$ below) is measured with respect to the rate of the normalisation decay $K^+\to\pi^+e^+e^-$ (denoted $K_{\pi ee}$ below), which allows a first order cancellation of detector and trigger inefficiencies. The $K_{\mu\nu ee}$ decay candidates are collected with the MT, $e$MT and $\mu$MT trigger lines, while the $K_{\pi ee}$ decay candidates are collected with the MT and $e$MT lines only. The following selection criteria are common for the $K_{\mu\nu ee}$ and $K_{\pi ee}$ decay candidates.
\begin{itemize}
\item Three-track vertices are reconstructed by extrapolating STRAW tracks into the FV, taking into account the measured residual magnetic field in the vacuum tank, and selecting triplets of tracks consistent with originating from the same point. Exactly one vertex should be present in the event. The total charge of the three tracks should be $q=+1$. The longitudinal position of the vertex, $z_{\rm vtx}$, should be within the FV. The momenta of the tracks forming the vertex should be in the range 6--44~GeV/$c$, and their trajectories through the STRAW chambers and extrapolated positions in the CHOD and LKr calorimeter front planes should be within the respective geometrical acceptances. Each pair of tracks should be separated by at least 15~(200)~mm in each STRAW chamber plane (LKr front plane) to suppress photon conversions and reduce shower overlap effects.
\vspace{-1mm}
\item Track times, $t_{\rm track}$, are defined initially using the CHOD information. The vertex CHOD time is evaluated as the average of the track CHOD times. Signals in the RICH geometrically compatible with the tracks, within 3~ns of the vertex CHOD time,
are used to evaluate track RICH times. Track and vertex time estimates are then refined using the more precise RICH information. Each track is required to be within 2.5~ns of the trigger time, $t_{\rm trigger}$.
\vspace{-1mm}
\item To suppress backgrounds with photons in the final state, originating from $K^+\to\pi^+\pi^0_D$ and $K^+\to\pi^0_D e^+\nu$ decays followed by the Dalitz decay $\pi^0_D\to\gamma e^+e^-$, no signals in the LAV detectors located downstream of the reconstructed vertex position are allowed within 4~ns of the vertex time.
\vspace{-1mm}
\item Particle identification is based on the ratio $E/p$ of the energy deposited in the LKr calorimeter (within 50~mm of the track impact point, within 10~ns of the vertex time) to the momentum measured by the spectrometer. Pion ($\pi^\pm$), muon ($\mu^\pm$) and electron ($e^\pm$) candidates are required to have $E/p<0.85$, $E/p<0.2$ and $0.9<E/p<1.1$, respectively. No geometrically associated MUV3 signals within 3~ns of the vertex time are allowed for pion candidates, and an associated MUV3 signal within 5~ns of the vertex time is required for the muon candidates.
\end{itemize}
The $K_{\pi ee}$ selection is identical to that of Ref.~\cite{co22}, and includes the following additional criteria.
\begin{itemize}
\item The tracks forming the vertex should be identified as $\pi^+e^+e^-$, according to the specified charge and particle identification requirements.
\vspace{-1mm}
\item The total momentum of the three tracks, $p_{\rm vtx}$, should satisfy the condition $|p_{\rm vtx}-p_{\rm beam}|<2~{\rm GeV}/c$, where $p_{\rm beam}$ is the beam central momentum. The total transverse momentum with respect to the beam axis should be below 30~MeV/$c$. The quantity $p_{\rm beam}$ and the beam axis direction are monitored throughout the data taking, typically every few hours,
with fully reconstructed $K^+\to\pi^+\pi^+\pi^-$ decays.
\vspace{-1mm}
\item The reconstructed $\pi^+e^+e^-$ mass, $m_{\pi ee}$, should be in the normalisation region defined as 470--505~MeV/$c^2$, accounting for the 1.7~MeV/$c^2$ resolution and the radiative tail. The reconstructed $e^+e^-$ mass should be $m_{ee}>140~{\rm MeV}/c^2$ to suppress backgrounds from the $K^+\to\pi^+\pi^0$ decay followed by $\pi^0_D\to e^+e^-\gamma$, $\pi^0_{DD}\to e^+e^-e^+e^-$, and $\pi^0\to e^+e^-$ decays. This leads to an acceptance reduction to 73\% of its value, i.e. a relative reduction of 27\%.
\end{itemize}
The following selection criteria are specific to the $K_{\mu\nu ee}$ selection. The presence of an undetected neutrino in the final state enhances the background, therefore the vertex position condition and the photon veto criteria are more stringent than in the $K_{\pi ee}$ case.
\begin{itemize}
\vspace{-0.8mm}
\item The tracks forming the vertex should be identified as $\mu^-e^+e^+$, according to the specified charge and particle identification requirements.
\vspace{-0.8mm}
\item A momentum deficit, $p_{\rm beam}-p_{\rm vtx}>10~{\rm GeV}/c$, is required to suppress the $K^+\to\pi^+\pi^+\pi^-$ background. This condition leads to a 55\% relative reduction of acceptance, assuming a uniform phase space distribution.
\vspace{-0.8mm}
\item The squared missing mass is defined as \mbox{$m_{\rm miss}^2\!=\!(P_K\!-\!P_\mu\!-\!P_{e1}\!-\!P_{e2})^2$}, where $P_K$, $P_\mu$ and $P_{e1,2}$ are the kaon, muon and positron four-momenta, respectively. The four-momenta are evaluated using the mean kaon beam momentum and the reconstructed daughter momenta, under the respective mass hypotheses. The signal region is defined as $-0.006~{\rm GeV}^2/c^4<m_{\rm miss}^2<0.004~{\rm GeV}^2/c^4$. The asymmetric definition reduces the $K^+\to\pi^+\pi^-e^+\nu$ background while maximising acceptance, taking into account the $m_{\rm miss}^2$ resolution of $1.4\times 10^{-3}$~GeV$^2$/$c^4$ and its non-gaussian tails.
\vspace{-0.8mm}
\item The longitudinal position of the vertex should not be within the first 3~m of the FV. This reduces the background from {\it upstream decays}, i.e. decays occurring upstream of GTK3. Track bending by the TRIM5 magnet leads to a biased reconstruction of the decay vertex and kinematic properties of these decays. 
\vspace{-0.8mm}
\item For further suppression of backgrounds with photons in the final state, no clusters in the LKr calorimeter are allowed with energy above 3~GeV, separated by more than 150~mm from each of the track impact points, and within 6~ns of the vertex time.
\vspace{-2mm}
\end{itemize}


\boldmath
\section{The effective number of $K^+$ decays}
\unboldmath
\label{sec:piee}

\begin{figure}[t]
\begin{center}
\resizebox{0.50\textwidth}{!}{\includegraphics{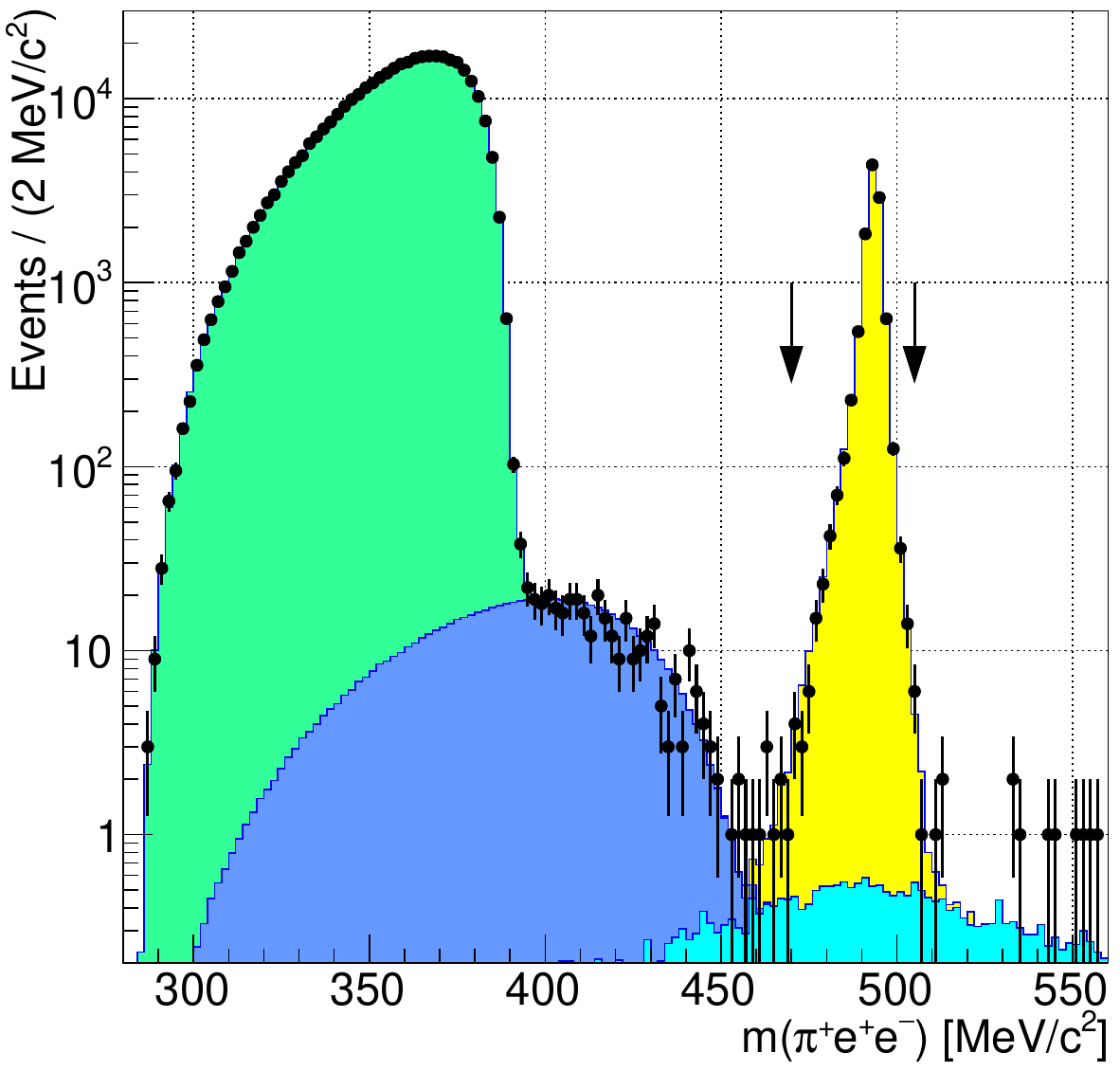}}%
\resizebox{0.50\textwidth}{!}{\includegraphics{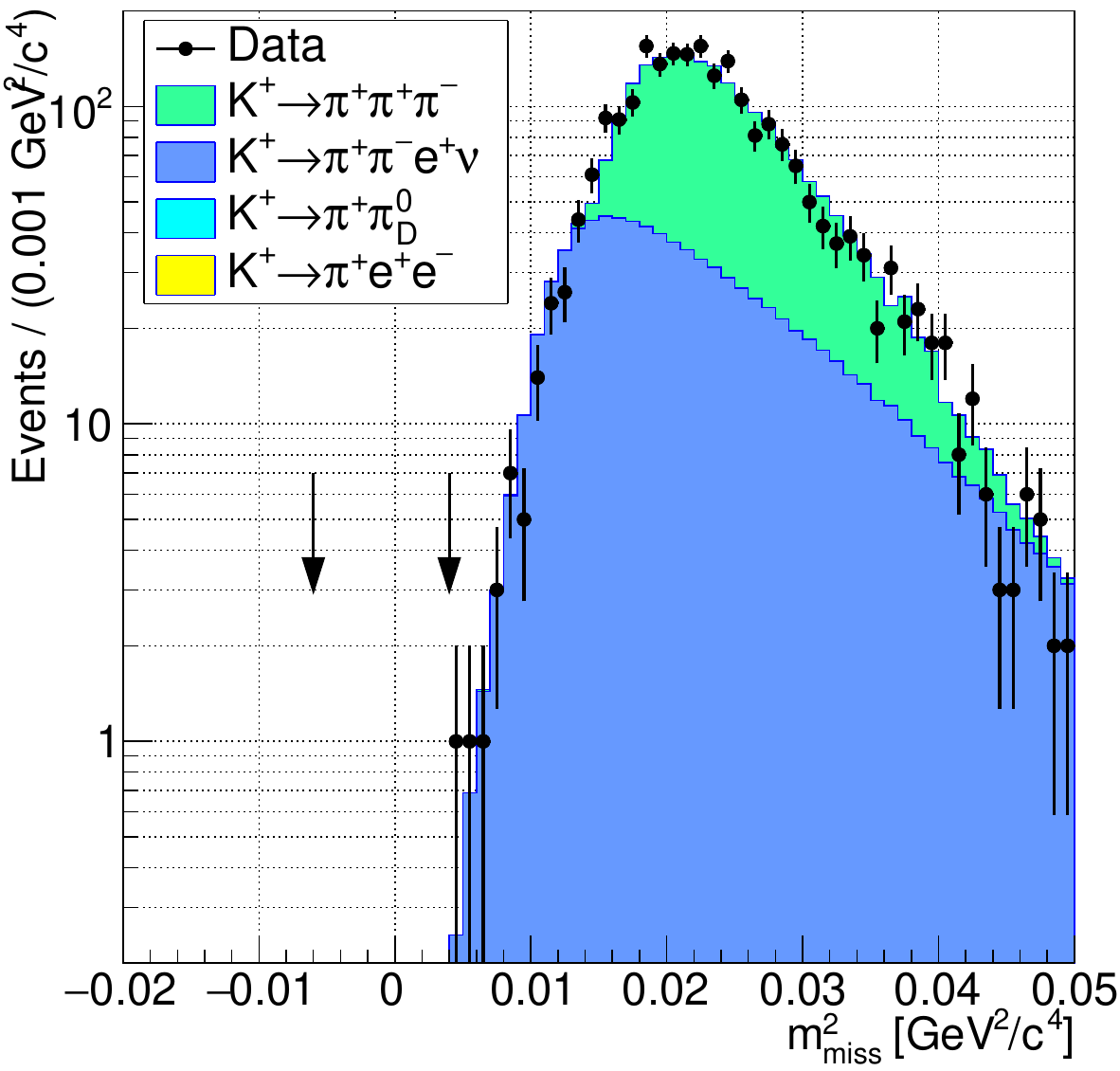}}
\end{center}
\vspace{-14mm}
\caption{Left: reconstructed $m_{\pi ee}$ spectra for the data and MC samples obtained with the $K_{\pi ee}$ selection. The data events are not weighted (see Section~\ref{sec:piee}). Right: reconstructed $m_{\rm miss}^2$ spectra for the data and MC samples obtained with the $K_{\mu\nu ee}$ selection. The normalisation and signal mass regions are indicated by vertical arrows.}
\label{fig:mass}
\end{figure}

The reconstructed $m_{\pi ee}$ spectra obtained with the $K_{\pi ee}$ selection for the data, as well as simulated signal and background components, are displayed in Fig.~\ref{fig:mass}~(left). Below the $m_{\pi ee}$ normalisation region, the background is mainly due to $K^+\to\pi^+\pi^+\pi^-$ decays with two pions ($\pi^\pm$) misidentified as electrons ($e^\pm$), and $K^+\to\pi^+\pi^-e^+\nu$ decays with one pion ($\pi^-$) misidentified as an electron ($e^-$). In the $m_{\pi ee}$ normalisation region, 10975 decay candidates are observed in the data sample, and the principal background comes from the $K^+\to\pi^+\pi^0_D$, $\pi^0_D\to\gamma e^+e^-$ decay chain. This background is suppressed by the $m_{ee}>140~{\rm MeV}/c^2$ selection condition, and contributes via double particle misidentification ($\pi^+\to e^+$ and $e^+\to\pi^+$). Pion and electron identification with the LKr calorimeter is modelled using (mis)identification probabilities measured from data samples of $K^+\to\pi^+\pi^+\pi^-$ and $K^+\to\pi^0 e^+\nu$ decays~\cite{co22}: the misidentification probabilities are about 1\%, and depend on momentum. Contribution to the background from the pion decay in flight, $\pi^\pm\to e^\pm\nu$, is negligible due to the ${\cal O}(10^{-4})$ branching fraction of this decay.

To account for the fact that the $\mu$MT trigger line is used to collect $K_{\mu\nu ee}$ events only, while the $e$MT and MT lines are used to collect both $K_{\mu\nu ee}$ and $K_{\pi ee}$ events, a weight determined by the trigger downscaling factors is applied to each $K_{\pi ee}$ event in the data sample to evaluate the number of $K_{\pi ee}$ candidates for normalisation:
\begin{displaymath}
w = \frac
{1-\left(1-\frac{1}{D_{e{\rm MT}}}\right)
\left(1-\frac{1}{D_{\mu{\rm MT}}}\right)
\left(1-\frac{1}{D_{{\rm MT}}}\right)}
{1-\left(1-\frac{1}{D_{e{\rm MT}}}\right)
\left(1-\frac{1}{D_{{\rm MT}}}\right)} \ge 1.
\end{displaymath}
The weight quantifies the enhancement of the kaon flux provided by the additional $\mu$MT trigger line used to collect $K_{\mu\nu ee}$ events. The weight has a typical value of 1.8, and reaches 2.9 for subsets of data with large values of the $D_{e\rm MT}/D_{\mu\rm MT}$ ratio.

The effective number of $K^+$ decays in the FV is computed as
\begin{displaymath}
N_K = \frac{(1-f)\cdot N_{\pi ee}}{{\cal B}_{\pi ee}\cdot A_{\pi ee}} = (1.97 \pm 0.02_{\rm stat} \pm 0.02_{\rm syst} \pm 0.06_{\rm ext})\times 10^{12},
\end{displaymath}
where: $N_{\pi ee}=21401$ is the number of weighted $K_{\pi ee}$ candidates in the data sample; ${\cal B}_{\pi ee}=(3.00\pm0.09)\times 10^{-7}$ is the $K_{\pi ee}$ branching fraction~\cite{pdg}; $A_{\pi ee}=(3.62\pm0.03_{\rm syst})\times 10^{-2}$ is the selection acceptance evaluated with simulations including trigger inefficiency and effects of event pileup; and $f=1.0\times 10^{-3}$ is the relative background contamination evaluated with simulations. The uncertainty in $A_{\pi ee}$ is estimated from stability checks with respect to variation of the selection criteria. The quoted systematic uncertainty in $N_K$ is due to $A_{\pi ee}$, while the external uncertainty is due to ${\cal B}_{\pi ee}$.


\boldmath
\section{Background to the $K^+\to\mu^-\nu e^+e^+$ decay}
\unboldmath
\label{sec:munuee}

\noindent {\bf Background due to single kaon decays}
\vspace{2mm}

\noindent Background to the $K_{\mu\nu ee}$ process from single $K^+$ decays is estimated using simulations with data-driven modelling of pion and electron (mis)identification, as described in Ref.~\cite{co22}. To validate the background estimates, {\it lower} and {\it upper} regions of $m_{\rm miss}^2$ located below and above the signal region are considered, while the signal region is kept masked.
\begin{itemize}
\item $K^+\to\pi^+\pi^+\pi^-$ decay, with double $\pi^+\to e^+$ misidentification and $\pi^-\to\mu^-\bar\nu$ decay in flight, contributes mainly in the upper $m_{\rm miss}^2$ region. Background in the signal mass region is minimised by the choice of the selection condition on the missing momentum, $p_{\rm beam}-p_{\rm vtx}$. Background from upstream $K^+\to\pi^+\pi^+\pi^-$ decays is minimised by the $z_{\rm vtx}$ selection condition.
\item $K^+\to\pi^+\pi^-e^+\nu$ decay, with $\pi^+\to e^+$ misidentification and $\pi^-\to\mu^-\bar\nu$ decay in flight, contributes mainly in the upper $m_{\rm miss}^2$ region. The contribution in the signal mass region is also minimised by the missing momentum and $z_{\rm vtx}$ selection conditions.
\item The $K^+\to\pi^0_D e^+\nu$, $\pi^0_D\to e^+e^-\gamma$ decay chain contributes via $e^-\to\mu^-$ misidentification if the photon is not detected. The electron misidentification probability achieved by the $E/p<0.2$ condition is found with simulations to be ${\cal O}(10^{-4})$, the probability of a matching in-time accidental MUV3 signal is ${\cal O}(10^{-2})$. Photon veto conditions provide further suppression, resulting in a small background contribution. The background from the rare decay $K^+\to e^+\nu e^+e^-$~\cite{bi93}, also contributing via $e^-\to\mu^-$ misidentification, is negligible.
\item The rare decay $K^+\to e^+\nu\mu^+\mu^-$~\cite{bi93} enters via muon decay in flight, $\mu^+\to e^+\nu\bar\nu$, and its contribution is small.
\end{itemize}

\vspace{1mm}
\noindent {\bf Background due to accidental activity}
\vspace{1.5mm}

\noindent Background due to coincidences of multiple kaon decays, beam pion decays or beam halo muons is estimated using several methods.
\begin{itemize}
\item Alternative event selections with an out-of-time track are considered: the timing condition $|t_{\rm track}-t_{\rm trigger}|<2.5~\rm{ns}$ is replaced by $2.5~{\rm ns}<|t_{\rm track}-t_{\rm trigger}|<20~{\rm ns}$ for either the $\mu^-$ candidate or one of the $e^+$ candidates. These selections are blind to single $K^+$ decays, and enhance accidental background by up to a factor of seven (taking trigger efficiencies into account). No data events in the signal $m_{\rm miss}^2$ region are observed using any of these alternative selections.
\vspace{-0.5mm}
\item Another alternative event selection is considered: the vertex charge condition is replaced by $q=-1$, and the $\mu^-e^+e^-$ final state is requested. No data events satisfy this selection in either of the signal, lower or upper regions of $m_{\rm miss}^2$.
\vspace{-0.5mm}
\item The background component due to the coincidence of two $K^+\to\pi^+\pi^+\pi^-$ decays is evaluated with a dedicated simulation: the estimated background in the signal $m_{\rm miss}^2$ region is $1.2\times 10^{-3}$ events.
\end{itemize}
It is concluded that background contributions due to accidental activity can be neglected.

\vspace{3mm}
\noindent {\bf Summary of background contributions}
\vspace{1.5mm}

\noindent The reconstructed $m_{\rm miss}^2$ spectra obtained with the $K_{\mu\nu ee}$ selection for the data, as well as simulated signal and background components, are displayed in Fig.~\ref{fig:mass}~(right). The estimated background contributions in the lower, signal and upper $m_{\rm miss}^2$ regions are listed in Table~\ref{tab:bkg}.
The numbers of data events in the lower and upper regions are compared to the background estimates before opening the masked region, and found to be in agreement within statistical fluctuations. The background in the signal region is estimated to be 
\begin{displaymath}
N_B = 0.26\pm 0.04,
\end{displaymath}
where the uncertainty is dominated by the MC statistical contribution.


\begin{table}[h]
\caption{Background estimates in the lower, signal and upper $K_{\mu\nu ee}$ squared missing mass regions with their statistical uncertainties. The contributions from upstream $K^+\to\pi^+\pi^+\pi^-$ and $K^+\to\pi^+\pi^-e^+\nu$ decays are quoted separately. Upper limits at 90\% CL are quoted when no simulated events satisfy the selection. The numbers of observed data events are also listed.}
\begin{center}
\vspace{-8mm}
\begin{tabular}{lrclrclrcl}
\hline
Mode / Region & \multicolumn{3}{c}{Lower} & \multicolumn{3}{c}{Signal} & \multicolumn{3}{c}{Upper} \\
\hline
$K^+\to\pi^+\pi^+\pi^-$ & \multicolumn{3}{c}{$<0.07$} & \multicolumn{3}{c}{$<0.07$} & 1412 & $\!\!\!\pm\!\!\!$ & 11 \\
$K^+\to\pi^+\pi^-e^+\nu$ & 0.01 & $\!\!\!\pm\!\!\!$ & 0.01 & 0.16 & $\!\!\!\pm\!\!\!$ & 0.02 & 867 & $\!\!\!\pm\!\!\!$ & 1 \\
$K^+\to\pi^+\pi^+\pi^-$ (upstream) & \multicolumn{3}{c}{$<0.03$} & 0.06 & $\!\!\!\pm\!\!\!$ & 0.03 & 1.5 & $\!\!\!\pm\!\!\!$ & 0.3 \\
$K^+\to\pi^+\pi^-e^+\nu$ (upstream) & 0.01 & $\!\!\!\pm\!\!\!$ & 0.01 & 0.01 & $\!\!\!\pm\!\!\!$ & 0.01 & 0.14 & $\!\!\!\pm\!\!\!$ & 0.03 \\
$K^+\to\pi^0_D e^+\nu$ & 0.02 & $\!\!\!\pm\!\!\!$ & 0.01 & 0.01 & $\!\!\!\pm\!\!\!$ & 0.01 & 0.02 & $\!\!\!\pm\!\!\!$ & 0.01 \\
$K^+\to e^+\nu\mu^+\mu^-$ & \multicolumn{3}{c}{$<0.01$} & \multicolumn{3}{c}{$<0.01$} & 0.05 & $\!\!\!\pm\!\!\!$ & 0.02 \\
\hline
Total expected & ~0.04 & $\!\!\!\pm\!\!\!$ & 0.02~ & ~0.26 & $\!\!\!\pm\!\!\!$ & 0.04~ & ~2281 & $\!\!\!\pm\!\!\!$ & 11~ \\
\hline
Data & \multicolumn{3}{c}{0} & \multicolumn{3}{c}{0} & \multicolumn{3}{c}{~~2271} \\
\hline
\end{tabular}
\end{center}
\vspace{-9mm}
\label{tab:bkg}
\end{table}


\vspace{-1mm}
\section{Results}
\vspace{-1mm}

The signal acceptance evaluated with simulations, assuming a uniform phase space distribution of signal events, is $A_{\mu\nu ee} = 0.0144$. The uncertainty in  $A_{\mu\nu ee}$ is negligible for the purpose of the signal search. The single event sensitivity, defined as the branching fraction of the $K_{\mu\nu ee}$ decay corresponding to the observation of one signal event, is found to be
\begin{displaymath}
{\cal B}_{\rm SES} = \left(N_K\cdot A_{\mu\nu ee}\right)^{-1} = (3.53\pm0.12)\times 10^{-11}.
\end{displaymath}
No data events are observed in the signal region after unmasking. An upper limit of the signal branching fraction is evaluated using the quantity ${\cal B}_{\rm SES}$ and the numbers of expected background events and observed data events using the CL$_{\rm S}$ method~\cite{re02}:
\begin{displaymath}
{\cal B}(K^+\to\mu^-\nu e^+e^+) < 8.1 \times 10^{-11} ~~ {\rm at} ~ 90\% ~ {\rm CL}.
\end{displaymath}


\section*{Summary}

A search for the forbidden decay $K^+\to\mu^-\nu e^+e^+$ has been performed using the dataset collected by the NA62 experiment at CERN in 2016--2018. An upper limit of $8.1\times 10^{-11}$ is obtained for the decay branching fraction at 90\% CL assuming a uniform phase space distribution of signal events, which improves by a factor of 250 over the previous search~\cite{di76,pdg}. The sensitivity is not limited by the background. Similarly to other limits for the rates of LNV/LFV decays, the result depends on the phase space density assumption. The sensitivity is not sufficient to obtain new constraints on the models involving Majorana neutrinos and lepton flavour violating ALPs and $Z'$, however the result probes physics beyond these models.


\section*{Acknowledgements}

It is a pleasure to express our appreciation to the staff of the CERN laboratory and the technical
staff of the participating laboratories and universities for their efforts in the operation of the
experiment and data processing.

The cost of the experiment and its auxiliary systems was supported by the funding agencies of 
the Collaboration Institutes. We are particularly indebted to: 
F.R.S.-FNRS (Fonds de la Recherche Scientifique - FNRS), under Grants No. 4.4512.10, 1.B.258.20, Belgium;
CECI (Consortium des Equipements de Calcul Intensif), funded by the Fonds de la Recherche Scientifique de Belgique (F.R.S.-FNRS) under Grant No. 2.5020.11 and by the Walloon Region, Belgium;
NSERC (Natural Sciences and Engineering Research Council), funding SAPPJ-2018-0017,  Canada;
MEYS (Ministry of Education, Youth and Sports) funding LM 2018104, Czech Republic;
BMBF (Bundesministerium f\"{u}r Bildung und Forschung) contracts 05H12UM5, 05H15UMCNA and 05H18UMCNA, Germany;
INFN  (Istituto Nazionale di Fisica Nucleare),  Italy;
MIUR (Ministero dell'Istruzione, dell'Universit\`a e della Ricerca),  Italy;
CONACyT  (Consejo Nacional de Ciencia y Tecnolog\'{i}a),  Mexico;
IFA (Institute of Atomic Physics) Romanian 
CERN-RO No. 1/16.03.2016 
and Nucleus Programme PN 19 06 01 04,  Romania;
INR-RAS (Institute for Nuclear Research of the Russian Academy of Sciences), Moscow, Russia; 
JINR (Joint Institute for Nuclear Research), Dubna, Russia; 
NRC (National Research Center)  ``Kurchatov Institute'' and MESRF (Ministry of Education and Science of the Russian Federation), Russia; 
MESRS  (Ministry of Education, Science, Research and Sport), Slovakia; 
CERN (European Organization for Nuclear Research), Switzerland; 
STFC (Science and Technology Facilities Council), United Kingdom;
NSF (National Science Foundation) Award Numbers 1506088 and 1806430,  U.S.A.;
ERC (European Research Council)  ``UniversaLepto'' advanced grant 268062, ``KaonLepton'' starting grant 336581, Europe.

Individuals have received support from:
Charles University Research Center (UNCE/SCI/ 013), Czech Republic;
Ministero dell'Istruzione, dell'Universit\`a e della Ricerca (MIUR \mbox{``Futuro} in ricerca 2012''  grant RBFR12JF2Z, Project GAP), Italy;
Russian Science Foundation (RSF 19-72-10096), Russia;
the Royal Society  (grants UF100308, UF0758946), United Kingdom;
STFC (Rutherford fellowships ST/J00412X/1, ST/M005798/1), United Kingdom;
ERC (grants 268062,  336581 and  starting grant 802836 ``AxScale'');
EU Horizon 2020 (Marie Sk\l{}odowska-Curie grants 701386, 754496, 842407, 893101, 101023808).

The data used in this paper were collected before February 2022.




\end{document}